\documentclass[reprint,superscriptaddress,amsmath,amssymb,aps,prm,floatfix]{revtex4-1}

\usepackage{graphicx}
\usepackage{dcolumn}
\usepackage{bm}
\usepackage{amsmath}
\usepackage{chemformula}[2014/04/08] \setchemformula{kroeger-vink} 
\usepackage{hyperref}
\usepackage{siunitx}
\usepackage{multirow}

\usepackage[normalem]{ulem}

\begin{document}


\title{Charged domain walls in improper ferroelectric hexagonal manganites and gallates}

\author{Didrik R. Sm{\aa}br{\aa}ten}
\affiliation{%
Department of Materials Science and Engineering, Faculty of Natural Sciences and Technology, NTNU Norwegian University of Science and Technology, NO-7491 Trondheim, Norway
}%
\author{Quintin N. Meier}
\affiliation{%
Materials Theory, ETH Zurich, Wolfgang-Pauli-Strasse 27, CH-8093 Zurich, Switzerland
}%
\author{Sandra H. Skj{\ae}rv{\o}}
\affiliation{%
Department of Materials Science and Engineering, Faculty of Natural Sciences and Technology, NTNU Norwegian University of Science and Technology, NO-7491 Trondheim, Norway
}%
\author{Katherine Inzani}
\affiliation{%
Department of Materials Science and Engineering, Faculty of Natural Sciences and Technology, NTNU Norwegian University of Science and Technology, NO-7491 Trondheim, Norway
}%
\author{Dennis Meier}
\affiliation{%
Department of Materials Science and Engineering, Faculty of Natural Sciences and Technology, NTNU Norwegian University of Science and Technology, NO-7491 Trondheim, Norway
}%
\author{Sverre M. Selbach}%
 \email{selbach@ntnu.no}
\affiliation{%
Department of Materials Science and Engineering, Faculty of Natural Sciences and Technology, NTNU Norwegian University of Science and Technology, NO-7491 Trondheim, Norway
}%

\date{\today}

\begin{abstract}
Ferroelectric domain walls are attracting broad attention as atomic-scale switches, diodes and mobile wires for next-generation nanoelectronics. Charged domain walls in improper ferroelectrics are particularly interesting as they offer multifunctional properties and an inherent stability not found in proper ferroelectrics. Here we study the energetics and structure of charged walls in improper ferroelectric YMnO$_3$, InMnO$_3$ and YGaO$_3$ by first principles calculations and phenomenological modeling. Positively and negatively charged walls are asymmetric in terms of local structure and width, reflecting that polarization is not the driving force for domain formation. The wall width scales with the amplitude of the primary structural order parameter and the coupling strength to the polarization. We introduce general rules for how to engineer $n$- and $p$-type domain wall conductivity based on the domain size, polarization and electronic band gap. This opens the possibility of fine-tuning the local transport properties and design $p$-$n$-junctions for domain wall-based nano-circuitry.
\end{abstract}

\maketitle
\section{INTRODUCTION}
Domain walls (DW) in ferroelectrics can be either charge neutral if the wall is oriented parallel to polarization ($P\parallel\mathrm{DW}$), or charged if the wall is oriented normal to the polarization ($P\perp\mathrm{DW}$). At charged walls, two polarization vectors point towards (head-to-head) or against (tail-to-tail) each other, leading to localized bound charges on the wall. This again leads to internal electric fields that drive an accumulation of mobile charge carriers and conducting 2D interfaces with great potential for nano-electronics. Additionally, these DW come with electric field configurations that are similar to the ones observed in $p$-$n$ junctions, foreshadowing the possibility to create DW-based transistors and logic gates.

Since the seminal observation of conducting ferroelectric DWs in BiFeO$_3$ \cite{PhysRevLett.105.197603,NatMater.8.229,PhysRevLett.107.127601}, DW functionality has been intensely studied in e.g. BaTiO$_3$\cite{NatCommun.4.1808}, LiNbO$_3$\cite{AdvFunctMater.22.3936}, Pb(Zr$_{0.2}$Ti$_{0.8}$)O$_3$\cite{AdvancedMaterials.23.5377} and hexagonal manganites (h-$R$MnO$_3$ where $R$=Y, In, Sc, Ho-Lu)\cite{PhysRevLett.108.077203,Nanotechnology.27.155705,NatMater.11.284,NatMater.16.622,PhysRevLett.118.036803,ApplPhysLett.104.232904}. The electrostatic energy cost of charged DWs makes them generally unstable in proper ferroelectrics\cite{NatCommun.4.1808}. In improper ferroelectrics, like h-$R$MnO$_3$, stable charged DWs occur naturally, because domain formation is not dominated by electrostatics, but by the critical dynamics of the non-polar primary mode.

Rare-earth hexagonal manganites, (h-$R$MnO$_3$), are type I multiferroics\cite{Nature.419.818}, with $T_C$ of $\sim$1250~K\cite{NatPhys.11.1070,PhysRevB.83.094111} and $T_N$ of $\sim$100~K\cite{Nature.419.818} depending on $R$. They undergo a geometrically driven cell-tripling improper ferroelectric phase transition\cite{NatMater.3.164,PhysRevB.72.100103} from $P6_3/mmc$ to $P6_3cm$\cite{PhysRevB.83.094111}. Condensation of the primary $K_3$ mode gives antiparallel displacements of $R$ along the z-axis (Fig.~\ref{fig:energetics_DFT}(a), and tilt of the Mn-O$_5$ trigonal bipyramids with a phase $\Phi$ and amplitude $Q$ \cite{PhysRevB.89.214107,NatMater.13.42}. As illustrated in Fig.~\ref{fig:energetics_DFT}(b-d), the two possible directions of polarization, and three different MnO$_5$ tilting directions, give rise to six different ferroelectric domain states ($\alpha^\pm,\beta^\pm,\gamma^\pm$) \cite{NatMater.9.253,NatMater.13.42}, which meet in a topologically protected vortex\cite{NatMater.9.253,PhysRevB.85.174422,PhysRevX.2.041022}. Charged DWs between these domains show suppressed or enhanced conductivity compared to bulk\cite{NatMater.11.284,NatMater.16.622,PhysRevLett.118.036803}, and the head-to-head DWs can be switched from resistive to conducting behavior by an applied electric field \cite{NatMater.16.622}. With their narrow width of $\sim$7~{\AA}\cite{NanoLett.17.5883}, these DWs are promising for DW-based circuitry.

YMnO$_3$ is the protoypical h-$R$MnO$_3$, while InMnO$_3$ has a similar band gap, but a smaller polarization\cite{PhysRevB.85.174422,ChemMater.29.2425,PhysRevB.87.184109}. Oppositely, YGaO$_3$ has a similar polarization to YMnO$_3$, but a larger band gap\cite{ActaCrystallogrB.31.2770,PhysRevB.79.144125}. These compounds cover a variation in polarization and band gap, which are important for conducting charged DWs.

Theoretical studies using density functional theory (DFT) have provided fundamental understanding of the physics of DWs in e.g. BiFeO$_3$\cite{NatMater.8.229,PhysRevB.80.104110,PhysRevLett.110.267601}, BaTiO$_3$\cite{PhysRevB.53.R5969}, PbTiO$_3$\cite{PhysRevB.68.134103,ApplPhysLett.75.2930}, and $R$MnO$_3$\cite{NatMater.11.284,NatCommun.4.1540}.
While neutral DWs in $R$MnO$_3$ have been studied by DFT\cite{NatCommun.4.1540} and atomistic simulations\cite{JPhysDApplPhys.48.435503}, charged DWs are more complex from a computational point of view due to the electrostatic potential across these walls.

Here we investigate the energetics, structure, and electronic properties of charged ferroelectric DWs in isostructural YMnO$_3$, InMnO$_3$ and YGaO$_3$ by phenomenological modeling and DFT calculations. We find that the energetics of different DW configurations agree with the Mexican hat energy landscape from Landau theory\cite{NatMater.13.42} and a trend in DW width with the evolution of the $K_3$ mode is found. A general criterion is suggested for when charged DWs become conducting. We hope our findings will guide future experiments and serve as a roadmap for engineering the properties of charged DWs for nano-electronic circuitry.\\

\section{RESULTS}
\begin{figure}
	\includegraphics[width=\linewidth]{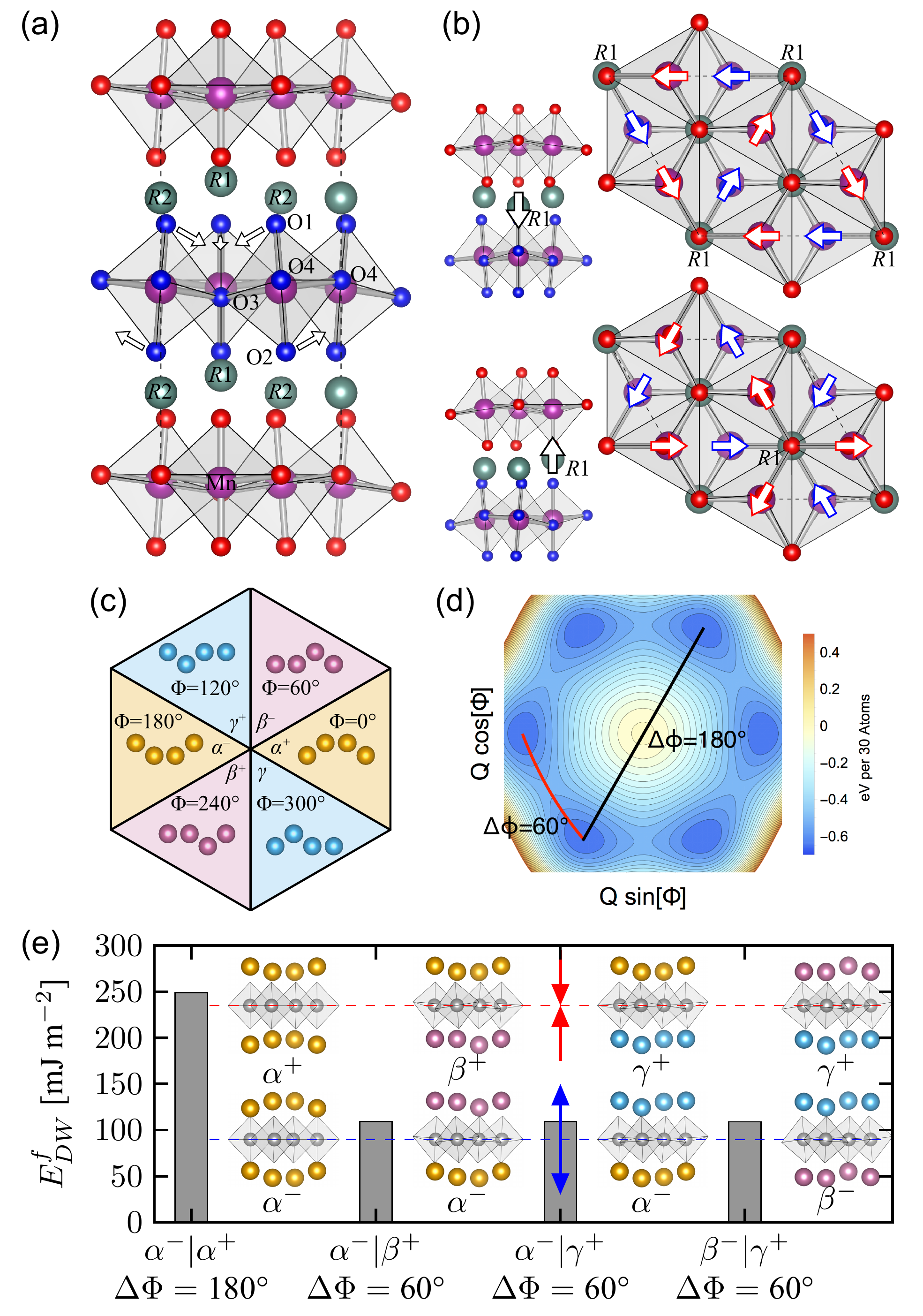}
	\caption{Energetics of charged domain walls. (a) Crystal structure of hexagonal $R$MnO$_3$ in the $P6_3cm$ space group, illustrating the $R$ cation corrugation and Mn-O$_5$ tilt. Changes in $\Phi$ with position of the trimerization center $R1$ and direction of cation is illustrated for $\alpha^+$ (top) and $\gamma^-$ (bottom) in (b). Red and blue arrows show the tilt direction of the trigonal bipyramids above and below the $R$ cation layer, respectively. (c) Sketch of topologically protected domain vortex surrounded by the six domains with respect to $R$ cation corrugation and $\Phi$, and (d) mapped energy landscape with respect to $(Q,\Phi)$, indicating the energy path for $\Delta\Phi=60^\circ$ or $\Delta\Phi=180^\circ$ DWs. (e) Calculated DW formation energies for $1\times1\times6$ supercells of YMnO$_3$ with $\Delta\Phi=180^\circ$ ($\alpha^-|\alpha^+$), and $\Delta\Phi=60^\circ$ ($\alpha^+|\beta^-$), ($\alpha^-|\gamma^+$), and ($\beta^-|\gamma^+$) configurations. Inset figures illustrate the resulting local structure at the head-to-head (top) and tail-to-tail (bottom) DW.}
	\label{fig:energetics_DFT}
\end{figure}

\begin{center}\textbf{A. Domain wall energetics.}\\
\end{center}
From symmetry constraints, there are two types of possible charged DWs in this system, one corresponding to a change of the phase of $\Delta\Phi=60^\circ$ and one corresponding to a change of phase of $\Delta\Phi=180^\circ$. As becomes clear from Fig.~\ref{fig:energetics_DFT}(d), for the $\Delta\Phi=180^\circ$ case, the transition path between the phases goes through the high-symmetry phase. This high energy transition path makes this kind of wall very costly, explaining why it has not been observed experimentally. The natural DWs in this material, occur with a change of phase of $\Delta\Phi=60^\circ$, which corresponds to a low-energy path in the energy landscape (Fig.~\ref{fig:energetics_DFT}(d). This is apparent from experimental results and symmetry constraints\cite{PhysRevB.89.214107,PhysRevLett.113.267602,PhysRevB.87.184109}, where all DWs within the material away from the vortices have $\Delta\Phi=60^\circ$.

Calculated DW formation energies for DFT relaxed $1\times1\times6$ supercells with three different $\Delta\Phi=60^\circ$ DW configurations, ($\alpha^-|\beta^+$), ($\alpha^-|\gamma^+$), and ($\beta^-|\gamma^+$), and one $\Delta\Phi=180^\circ$ configuration, ($\alpha^-|\alpha^+$), are compared in Fig.~\ref{fig:energetics_DFT}(e) (vertical line $|$ represents a DW). The three $\Delta\Phi=60^\circ$ configurations are degenerate within 0.03 meV/atom, and $\sim$140 mJ m$^{-2}$ lower in energy than the $\Delta\Phi=180^\circ$ configuration. The large energy difference between $\Delta\Phi=180^\circ$ and $\Delta\Phi=60^\circ$ can be explained by the aforementioned energy landscape (Fig~\ref{fig:energetics_DFT}(d), as well as the symmetry breaking and the distinct structural changes across the walls (Fig. S1 \cite{makeref4SI}). These findings are consistent with experimental TEM studies\cite{NanoLett.17.5883}, and are important for switching dynamics as high energy DWs can only occur transiently. We note that due to the electrostatic potential across the cell model, the DW energy is inherently cell size dependent as discussed later.\\
 
\begin{center}\textbf{B. Structural evolution across DWs.}\\ 
\end{center}
From now on we will concentrate on the naturally occurring $\Delta\Phi=60^\circ$ DWs. The energy landscape of the involved phases can be described by the following Landau free energy functional\cite{NatMater.13.42}:
\begin{align*}
	F[Q,\Phi,P]&=\dfrac{a}{2}Q^2+\dfrac{b}{4}Q^4+\dfrac{Q^6}{6}(c+c'\cos6\Phi)\\
	&-gQP^3\cos3\Phi+\dfrac{g'}{2}P^2Q^2+\dfrac{a_p}{2} P^2\\
    &+\sum\limits_{i\in x,y,z}\dfrac{s_Q^i}{2}\left[(\partial_i Q)^2+Q^2(\partial_i\Phi)^2\right]+\dfrac{s^{i}_{p}}{2}(\partial_iP)^2\,.
\end{align*}
Here the order parameter amplitude $Q$ and phase $\Phi$ describe the amplitude and angle of the tilt as shown in Fig.~\ref{fig:energetics_DFT}(a-b), and $P$ is the spontaneous polarization, which is coupled non-linearly to $Q$. The formation of DWs within this formalism can be studied by minimizing this free energy with fixed boundary conditions. The evolution of the amplitudes of the different involved modes across a DW can thus be predicted by minimizing the free energy functional with boundary conditions $\Phi=0$ at $x=-\infty$ and $\Phi=2\pi/6$ at $x=\infty$. The DW width is defined by the competition of the energy landscape, where the DW becomes as narrow as possible, while the gradient terms lead to a broadening. The DW width is quantified by the evolution of $\Phi$ through the approximate analytical solution derived by Holtz \textit{et al.}\cite{NanoLett.17.5883}:
\begin{equation}\label{eq:dw width landau}
	\Phi(z)=\Phi_n+\dfrac{2}{3}\arctan(e^{z/\xi_6})\quad ,
\end{equation}
where $\Phi_n$ is the phase angle for domain $n$, and $\xi_6$ the characteristic length associated to the DW width.

\begin{figure}
	\includegraphics[width=\linewidth]{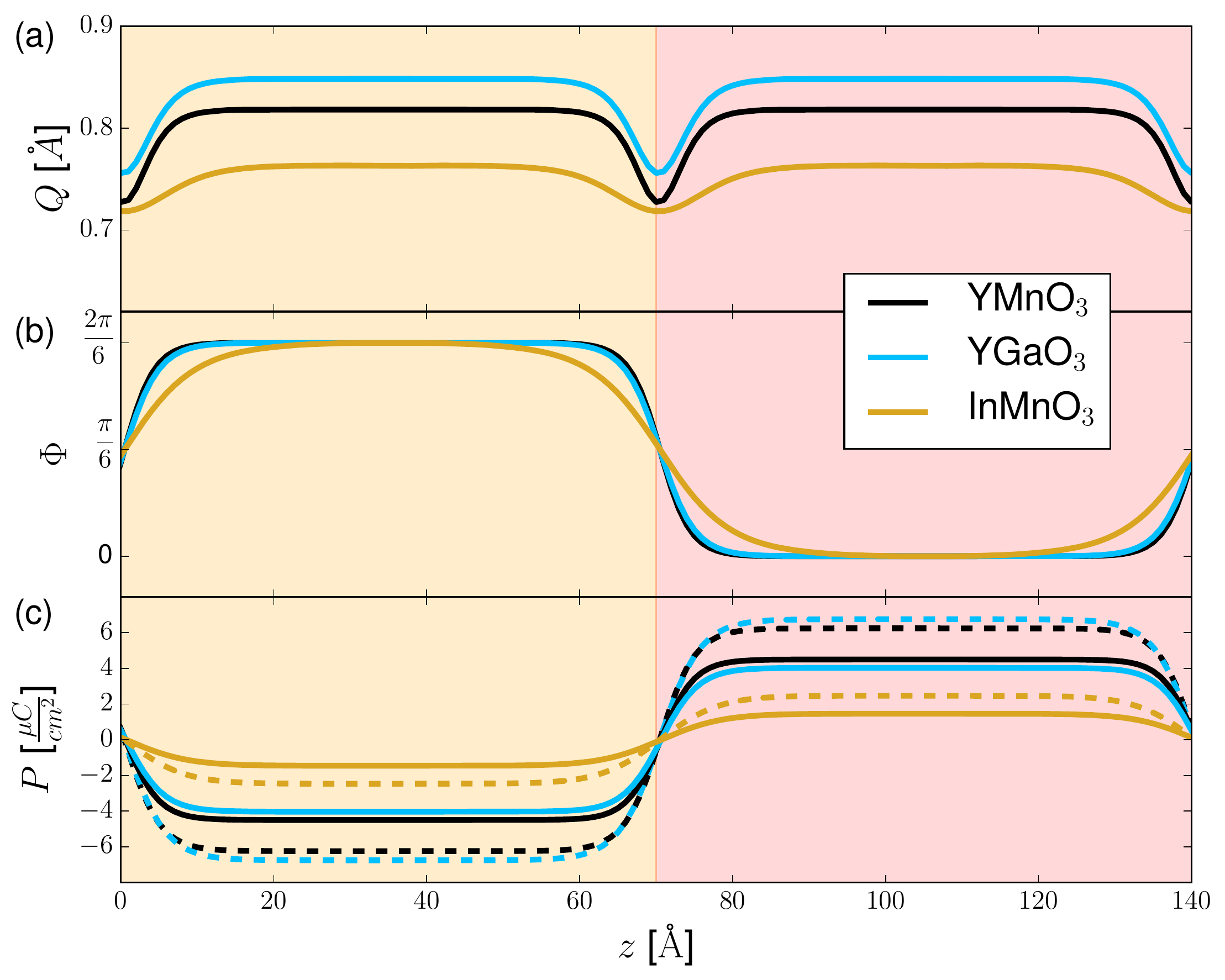}
	\caption{Crystal structure evolution. Mode amplitudes across Ferroelectric DWs in YMnO$_3$, YGaO$_3$ and InMnO$_3$, (a) trimerization amplitude $Q$, (b) trimerization phase $\Phi$ and (c) polarization $P$ with ($-$) and without ($--$) dielectric screening of the depolarizing field.}
	\label{fig:struct_Landau}
\end{figure}

The free energy is numerically minimized for the three compounds extracted from first principles calculations (see details in \cite{makeref4SI}). The amplitudes of the primary order parameter ($Q$,$\Phi$) and secondary order parameter $P$ are shown in Fig~\ref{fig:struct_Landau}(a-c). These predictions are fully consistent with experimental TEM observations\cite{NanoLett.17.5883}. In general, the DWs in YGaO$_3$ and YMnO$_3$ are very similar, while the DWs in InMnO$_3$ are significantly broader. The main reason for this is the inherently smaller amplitude of $Q$ and the weaker coupling term to $P$ in this material, reasoned from the more covalent nature of the In-O bond compared to the more ionic Y-O bond\cite{PhysRevB.85.174422}. The energy cost of forming DWs is expected to be significantly lower for InMnO$_3$ compared to YMnO$_3$ and YGaO$_3$, as apparent from the calculated DW formation energies in Table~\ref{tab:1}. In addition, we show in Fig.~\ref{fig:struct_Landau}(c) the difference of the polarization for screened and non-screened electric fields, which we will discuss more later on. We list calculated DW widths in Table~\ref{tab:1}.\\

\begin{table}
\caption{\label{tab:1}Calculated domain wall width, $\xi_6$, in $1\times1\times6$ $\Delta\Phi=60^\circ$ ($\alpha^-|\beta^+$) supercell for head-to-head (h-t-h) and tail-to-tail (t-t-t) DW from DFT calculations and from Landau theory. DW formation energy $E^f_{DW}$ in $1\times1\times6$ $\Delta\Phi=60^\circ$ ($\alpha^-|\beta^+$) supercell, band gap $E_g$ (from DFT), and calculated background ($\epsilon_b$) and ground state ($\epsilon_{tot}$) dielectric constants along c. }
\begin{ruledtabular}
\begin{tabular}{lccc}
	Property & YMnO$_3$ & InMnO$_3$ & YGaO$_3$ \\ \hline
 $\xi_6$, Landau [{\AA}] & 2.4    & 5.7   & 2.8  \\
 $\xi_6$, DFT, h-t-h [{\AA}]  & 1.7   & 6.8  & 3.0 \\
 $\xi_6$, DFT, t-t-t [{\AA}]  & 2.2   & 6.8  & 3.2 \\
 $E^f_{DW}$, DFT, [mJ~m$^{-2}$]    & 109.4 & 22.6 & 115.0 \\
 $Eg$, DFT [eV] & 1.5 & 1.4 & 3.1\\
 $\epsilon_b$ & 8.9 & 10.3 & 6.7 \\
 $\epsilon_{tot}$ & 12.5 & 14.8 & 11.2 \\
\end{tabular}
\end{ruledtabular}
\end{table}

\begin{figure}
	\includegraphics[width=\linewidth]{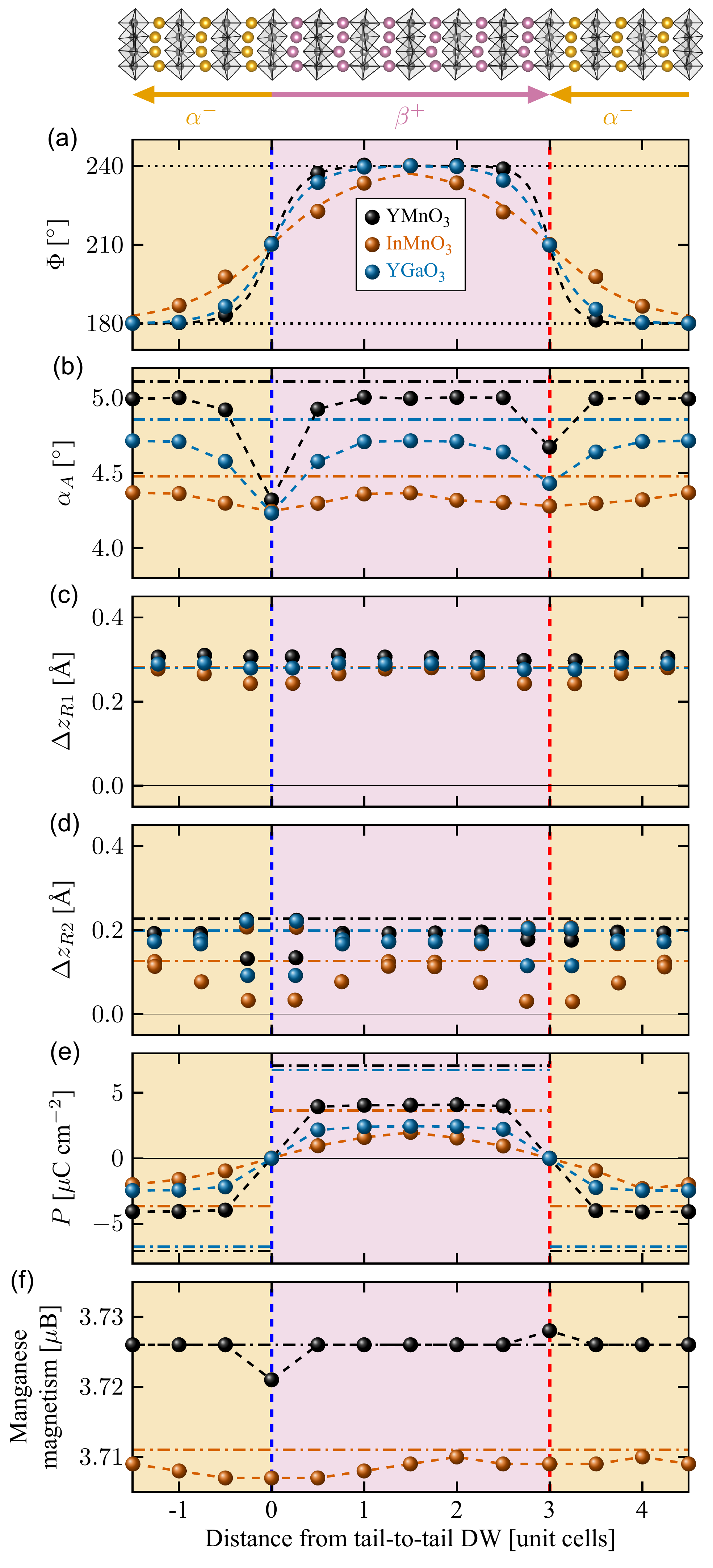}
	\caption{Local crystal structure across charged DWs. Layer resolved (a) phase $\Phi$ with fitted profile from Eq. \eqref{eq:dw width landau}, (b) amplitude $\alpha_A$, $R$ cation displacements (c) $\Delta z_{R1}$ and (d) $\Delta z_{R2}$, and (e) polarization $P$ across YMnO$_3$, InMnO$_3$, and YGaO$_3$ $1\times1\times6$ supercells with $\Delta\Phi=\SI{60}{\degree}$ ($\alpha^-|\beta^+$) configuration. (f) Magnetic moments of Mn in YMnO$_3$ and InMnO$_3$. Dash-dotted horizontal lines in (a-f) show bulk $\Phi$, $\alpha_A$, $\Delta z_{Ri}$, $P$, and Mn magnetism, from DFT relaxed 30 atom unit cells. Dashed vertical lines illustrate the position of the tail-to-tail (left) and head-to-head (right) DWs. Yellow shading represents $P_{\downarrow}$ domain ($\alpha^-$), pink shading $P_{\uparrow}$ domain ($\beta^+$).}
	\label{fig:struct_DFT}
\end{figure}

\begin{center}\textbf{C. Local crystal structure.}\\
\end{center}Using $1\times1\times6$ supercells with $\Delta\Phi=60^\circ$ ($\alpha^-|\beta^+$) configuration as our DFT model, we next address the local structural changes across the DWs in terms of phase $\Phi$, amplitude $Q$, $\alpha_A$, $R$ cation displacements $\Delta z_{Ri}$, and polarization $P$, Fig.~\ref{fig:struct_DFT}. $Q$ is here represented by the tilt angle of the apical oxygen relative to the $c$ axis, $\alpha_A$, according to Skj{\ae}rv{\o} \textit{et al.}\cite{skjaervo_unconventional_2017}.

The three compounds behave differently with respect to $\Phi$ (Fig.~\ref{fig:struct_DFT}(a); YMnO$_3$ shows the most abrupt change across the walls, YGaO$_3$ is intermediate, while InMnO$_3$ shows the most gradual change. From the $\Phi$ profiles fitted to Eq. \eqref{eq:dw width landau}, YMnO$_3$ has the most narrow DWs of $\sim$2~{\AA}, YGaO$_3$ intermediate with $\sim$3~{\AA}, and InMnO$_3$ the widest DWs of $\sim$7~{\AA}.
The same trend can be seen from the $\alpha_A$ profiles. In YMnO$_3$, a step-like change in $\alpha_A$ is observed across the head-to-head wall, while across the tail-to-tail wall a smoother $\alpha_A$ evolution is observed. YGaO$_3$ shows smoother $\alpha_A$ evolution across both DWs compared to the more step-like $\alpha_A$ profile in YMnO$_3$, while InMnO$_3$ shows an almost flat $\alpha_A$ profile.

Comparing the numerically minimized continuum picture with fully relaxed DWs (Table~\ref{tab:1}), it is apparent that while the results agree quantitatively, there are several subtleties in the atomic structure. We also see that the supercell we use for InMnO$_3$ in our DFT calculations can barely fit the two DWs, in line with the experimental observations\cite{PhysRevLett.113.267602}.

$R1$ cation displacements show little to no change across the supercell for all three compounds, Fig. \ref{fig:struct_DFT}(c). In contrast, $R2$ cations show a much greater variation across the supercell. This is due to the strong coupling between $R1$ and the closest planar O3 (2.29~{\AA} in YMnO$_3$) compared to the weaker $R2$-O4 bond (2.41~{\AA} in YMnO$_3$)\cite{skjaervo_unconventional_2017}. Close to the DWs the two $R2$ cations in each layer become non-equivalent with respect to $z$-component, with one of the two $R2$ shifting closer to the high symmetry position, while the other $R2$ shifts towards the $\Delta z_{R1}$ value (Fig. \ref{fig:struct_DFT}(d). The resulting up-intermediate-down $R$ cation corrugation resembles that of the centrosymmetric anti-polar phase, which has been found in InMnO$_3$ \cite{NatMater.13.42,PhysRevB.85.174422,PhysRevLett.113.267602,ChemMater.29.2425}.

Interestingly, the head-to-head and tail-to-tail DWs show asymmetric crystal structures, where the latter is wider. Looking closer at the local $R$-O$_7$ chemical environment at the walls (Fig. S2 \cite{makeref4SI}), head-to-head will be $R2$ terminated, while the tail-to-tail will be $R1$ terminated. Because of the strong $R1$-O3 bond described above, the local symmetry at the DW centre at the head-to-head wall will be structurally screened by the $R1$-O3 bonding to the neighbouring Mn-O$_5$ layer, resulting in narrower walls. Oppositely, at the tail-to-tail wall, the strong $R1$-O3 bond is towards the DW center. The weaker $R2$-O4 bond to the neighboring Mn-O$_5$ layer will not structurally screen the wall. $R1$-O3 is observed first between neighboring Mn-O$_5$ and the next-neighbouring $R$-cation layer, resulting in wider walls. The local chemical environment is addressed in more detail in \cite{makeref4SI}.\\

\section{DISCUSSION}
\textbf{Local electric fields.}
Head-to-head and tail-to-tail walls are of high interest because gradients in the polarization lead to bound charge on the interface, creating internal electric fields in the material. The induced electric field $\mathcal{E}$ is described by Gauss' Law:
\begin{equation}\label{gauss}
	\nabla\cdot\vec{\mathcal{E}}=-\dfrac{\nabla\cdot\vec{P}}{\epsilon_b\epsilon_0} \quad ,
\end{equation}
where $P$ is the spontaneous polarization, while the electronic screening and the the contribution of the other normal modes are described by the background dielectric constant $\epsilon_b$.
The electric field created by the bound charge on a straight head-to-head or tail-to-tail DW is given by:
\begin{equation}\label{efield}
	\vec{\mathcal{E}}=-\dfrac{\Delta P}{2 \epsilon_b\epsilon_0} \quad ,
\end{equation}
where $\Delta P$ is the change of spontaneous polarization across the DW. The amplitude of the spontaneous polarization $P$ should not be understood as being fixed to bulk value in this formula, since the polar mode itself has a high contribution to the dielectric screening in the material\cite{PhysRevB.86.094112}.

\begin{figure*}
	\includegraphics[width=\linewidth]{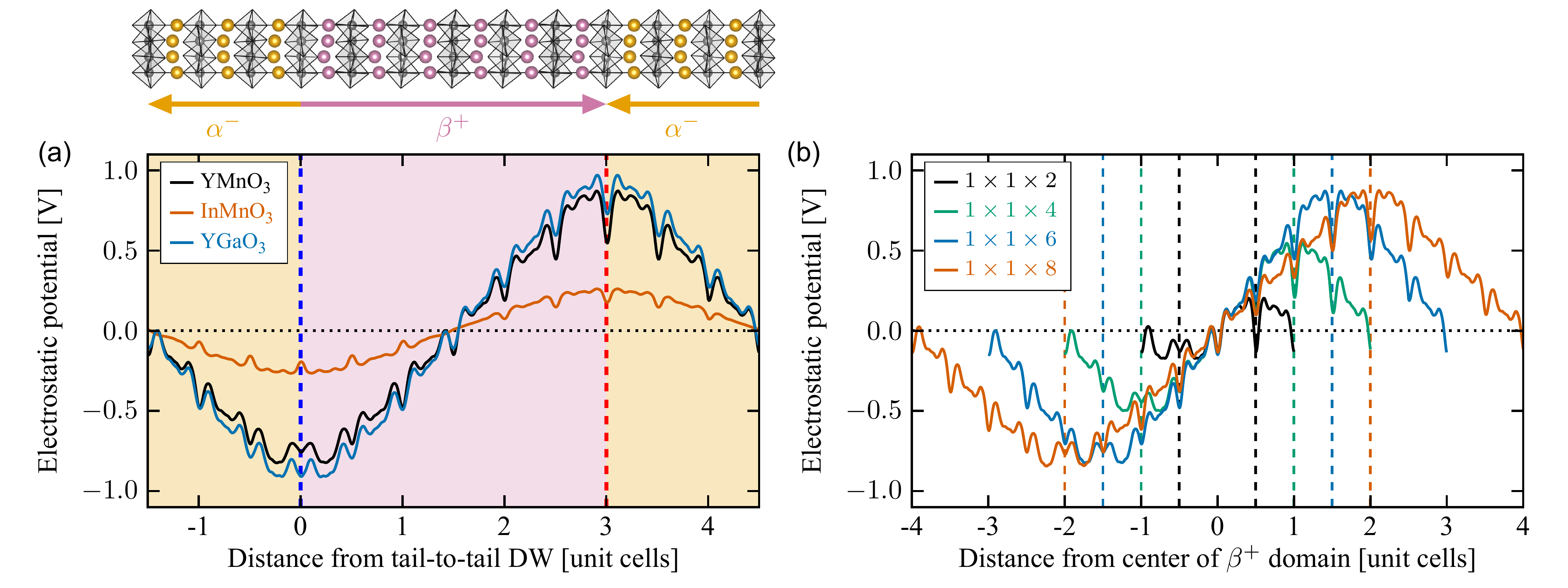}
	\caption{Electrostatic potential. (a) Macroscopically averaged electrostatic potential across $1\times1\times6$ $\Delta\Phi=60^\circ$ ($\alpha^-|\beta^+$) supercells of YMnO$_3$ (black), InMnO$_3$ (red) and YGaO$_3$ (blue). (b) Electrostatic potential for increasing DW distance in $\Delta\Phi=60^\circ$ ($\alpha^-|\beta^+$) YMnO$_3$. The dashed vertical lines in b) represent the position of the tail-to-tail (left) and head-to-head (right) walls for the different supercell sizes.}
	\label{fig:elstat_DFT}
\end{figure*}

The total field also depends on the topography of the domain structure and if the number of stacked walls is even or odd (See Fig. S3 in \cite{makeref4SI}). For simplicity we here address a periodic array of alternating walls identical to the case studied by DFT. The macroscopically averaged electrostatic potentials from DFT relaxed $1\times1\times6$ supercells are shown in (Fig.~\ref{fig:elstat_DFT}(a). The tail-to-tail wall is negative and head-to-head positive, in accordance with an electric field described by Eq.~\eqref{efield}. The electrostatic potential in YMnO$_3$ (Fig.~\ref{fig:elstat_DFT}(b) is independent of DW distance up until a critical distance, as expected for the field obtained from sheets of charge.
The existence of these electric fields leads to a dielectric response of the material in the form of reduced spontaneous polarization.

The resulting electric field can be calculated from the extended free energy with an additional contribution from the electrostatic energy:
\begin{equation}\label{Ftot}
	F_{tot}[Q,\Phi,P]=F[Q,\Phi,P]-P\mathcal{E} \quad ,
\end{equation}
where $\mathcal{E}$ is calculated using \eqref{gauss}.
The effect of this inherent depolarizing field is found by self-consistently solving \eqref{Ftot} and \eqref{gauss}, where $\mathcal{E}$ is recalculated after each minimization step in \eqref{Ftot}.
The substantial reduction in polarization within the domains due to this arising depolarizing field is shown in Fig.~\ref{fig:struct_Landau}. These phenomenological predictions of the resulting polarization compare well to what we observe from relaxed DFT, where the trimerization ($Q,\Phi$) is close to bulk values inside the domains (Fig.~\ref{fig:struct_DFT}(a-b), while $P$ is strongly reduced compared to bulk (Fig.~\ref{fig:struct_DFT}(e) and Table~\ref{tab:2}).\\

\begin{table}
\caption{\label{tab:2}Calculated polarization $P$ from DFT relaxed 30 atom unit cells, and from center of bulk $1\times1\times6$ $\Delta\Phi=60^\circ$ ($\alpha^-|\beta^+$) by point charge model. Calculated $P$ from Landau model by minimizing the Free Energy with respect to $K_3$ and $\Gamma_2$-modes, and by introducing electrostatics.}
\begin{ruledtabular}
\begin{tabular}{lcccc}
	\multirow{2}{*}{System} & \multicolumn{2}{c}{$P$, DFT [$\mu$C~cm$^{-2}$]} & \multicolumn{2}{c}{$P$, Landau [$\mu$C~cm$^{-2}$]} \\
	& Unit cell & $1\times1\times6$ & $K_3+\Gamma_2$ & w/El.statics \\ \hline
 YMnO$_3$  & 7.05 & 4.07 & 6.3 & 4.5 \\
 InMnO$_3$ & 3.64 & 1.74 & 2.5 & 1.5\\
 YGaO$_3$  & 6.72 & 2.42 & 6.8 & 3.9\\
\end{tabular}
\end{ruledtabular}
\end{table}

\textbf{Band bending and electrostatic breakdown.} The electric field induced by the walls is independent of DW distance (Fig.~\ref{fig:elstat_DFT}(b) if the domains are sufficiently small. Above this domain size, or DW distance, the electric potential becomes too large, leading to a rearrangement of charge in the system. Since both cases are interesting for potential applications, such as transistors and conducting sheets or channels, we address the screening as a function of domain size -- the distance between the DWs. This is particularly interesting in the h-$R$MnO$_3$ as the domain size can be controlled by the cooling rate through $T_C$\cite{PhysRevX.2.041022,Meier.PhysRevX.7.041014}.

In the presence of the electric fields, the energy of the valence and the conduction bands are given by:
\begin{equation}
	\begin{split}
		E_\text{VB,bent} & =E_\text{VB}+e^-\mathcal{E}d \quad , \\
		E_\text{CB,bent} & =E_\text{CB}+e^-\mathcal{E}d \quad .
	\end{split}
\end{equation}

This leads to a Zener-like breakdown if
\begin{equation}\label{eq:charge_transfer}
	E_g=e^-\mathcal{E}d \quad ,
\end{equation}
where $E_g$ is the band gap energy. In an infinite array of DWs, we find that $\mathcal{E}={(\epsilon\epsilon_0)^{-1}}{P}$ (see Fig. S3) and that charge is transferred from one kind of wall to the other when this breakdown criterion is fulfilled. This leads to the critical distance between the walls of
\begin{equation}\label{dbreakdown}
d=\dfrac{\epsilon_0\epsilon E_g}{e^{-}P} \quad .
\end{equation}

When the distance between DWs exceeds this critical distance, the system will transfer charges from one wall to another. This will give local occupation of the conduction band at the tail-to-tail wall, and correspondingly holes in the valence band at the head-to-head wall, leading to local $n$-type and $p$-type conductivity, respectively. The band bending can be observed as a change of the Fermi energy in the local electronic density of states at the DWs. When the distance between two DWs exceeds the critical distance $d$, the Fermi level at the DWs dips into the conduction band at the head-to-head wall, and into the valence band at the tail-to-tail wall. 

The electrostatic potential profile will flatten out close to the interfaces. This is apparent for the electrostatic potential gradients with increasing supercell size (Fig~\ref{fig:elstat_DFT}(b), where the electrostatic potential tends to gradually flatten out with increasing DW distance.

\begin{figure}
	\includegraphics[width=\linewidth]{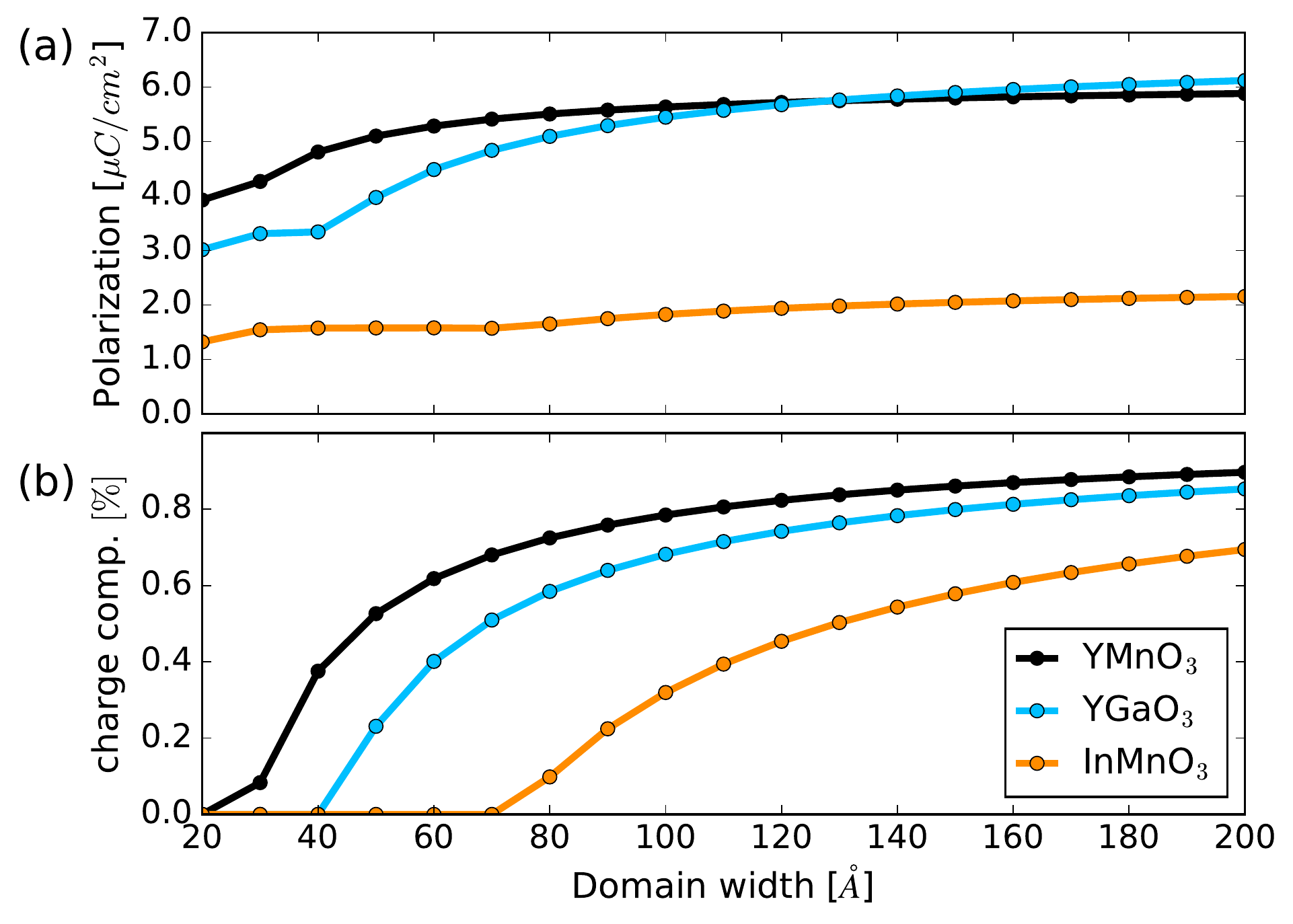}
	\caption{Electrostatic breakdown. (a) Polarization and (b) charge compensation at the walls as a function of domain width for YMnO$_3$, InMnO$_3$, and YGaO$_3$. We note that the kink in YGaO$_3$ is due to the saturation of the polar mode due to small domain size versus the increase due to screening. DFT calculated band gaps were used in the calculations.}
	\label{fig:charge_comp_Landau}
\end{figure}

To calculate the amount of charge compensation on the different DWs, the Landau model is extended: when the potential is higher than the band-gap, the system is allowed to rearrange charges. From this model, we can extract a polarization vs. domain width, and polarization vs. amount of free charge on the wall, shown in Fig.~\ref{fig:charge_comp_Landau}. The $\mathcal{E}$-field decreases steadily the broader the domain becomes up to several hundred \AA.

\begin{figure}
	\includegraphics[width=\linewidth]{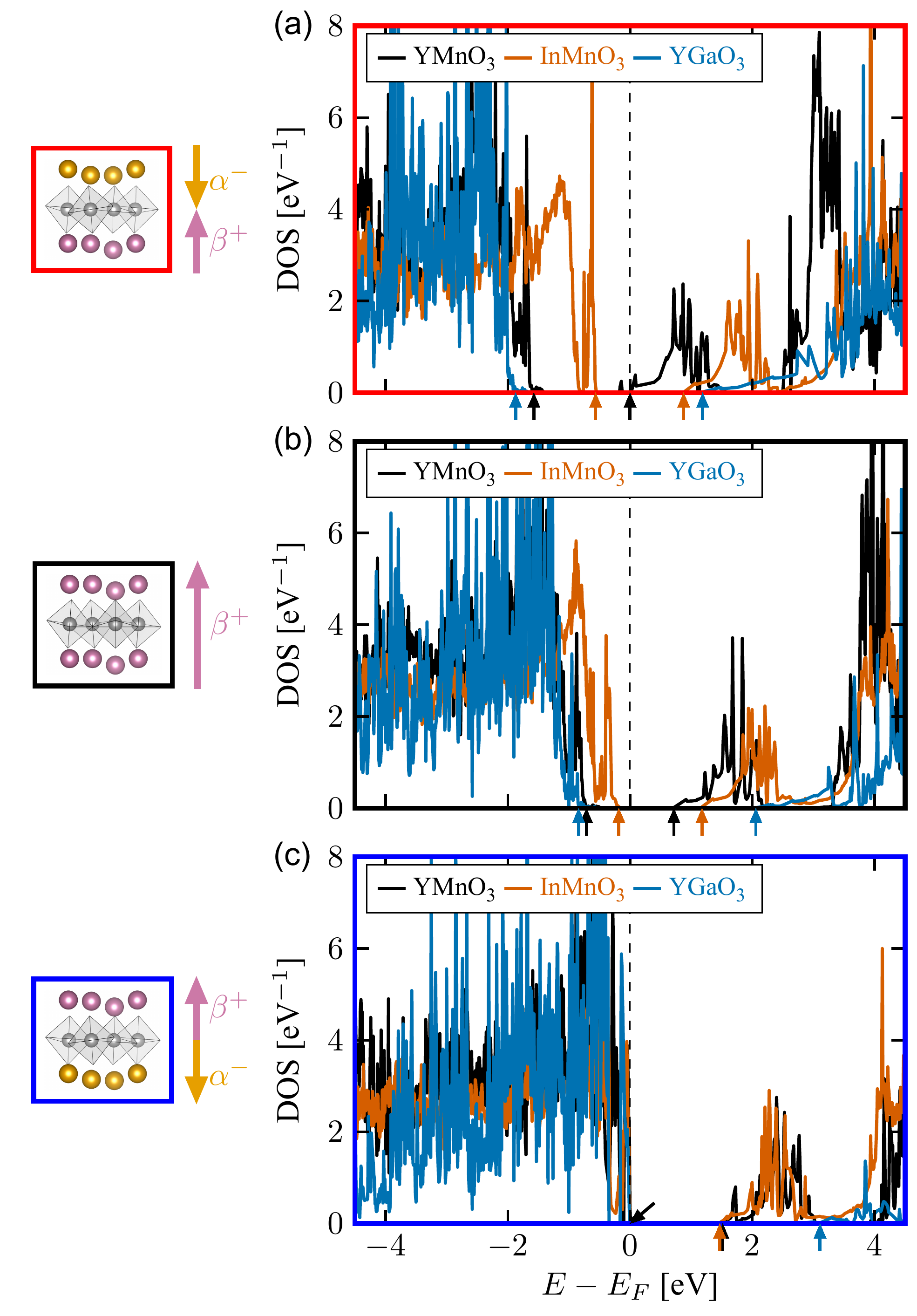}
	\caption{Local electronic structure. Calculated local electronic density of states for $1\times1\times6$ $\Delta\Phi=60^\circ$ ($\alpha^-|\beta^+$) supercells of YMnO$_3$, InMnO$_3$, and YGaO$_3$, illustrating the change in the Fermi level position across the supercell. The local DOS is shown (a) at the head-to-head wall, (b) at the centre of the $\beta^+$ domain, and (c) at the tail-to-tail wall. Arrows indicate the position of the valence band maximum and conduction band minimum.}
	\label{fig:DOS_DFT}
\end{figure}

The calculated local electronic density of states at a head-to-head wall, in the centre of a $\beta^+$ domain, and at a tail-to-tail wall in $1\times1\times6$, $\Delta\Phi=60^\circ$ ($\alpha^-|\beta^+$) supercells of YMnO$_3$, InMnO$_3$, and YGaO$_3$ are shown in Fig.~\ref{fig:DOS_DFT}. 
At the head-to-head wall in YMnO$_3$, the Fermi level is located above the conduction band edge, indicating $n$-type behavior. Oppositely, at the tail-to-tail wall, unoccupied states at the valence band edge are observed, indicating $p$-type behavior. Insulating behavior is predicted for bulk YMnO$_3$, in agreement with previous work\cite{NatMater.11.284}. 

This screening of the bound positive and negative charges at the head-to-head and tail-to-tail DWs is expected to be realized by partial reduction of Mn$^{3+}$ to Mn$^{2+}$ and oxidation of Mn$^{3+}$ to Mn$^{4+}$, respectively. While Bader charge analysis did not give clear trends (Fig. S4 \cite{makeref4SI}), the planar averaged Mn magnetic moments in Fig.~\ref{fig:struct_DFT}(f) follow a stepwise increase from 3.726~$\mu$B in bulk to 3.728~$\mu$B at the head-to-head DW centre, and a stepwise reduction to 3.721~$\mu$B at the tail-to-tail DW centre. This indicates subtle partial reduction and oxidation of manganese, respectively. It can also be seen clearly in Fig.~\ref{fig:DOS_DFT}, where for YMnO$_3$ the Fermi level moves into the lowest unoccupied Mn-d-states at the head-to-head wall, and into the highest occupied Mn-d-states at the tail-to-tail wall.

Compared to YMnO$_3$, neither InMnO$_3$ nor YGaO$_3$ show charged walls, Fig.~\ref{fig:DOS_DFT}. YMnO$_3$ and YGaO$_3$ have similar polarization, but YGaO$_3$ has a significantly higher band gap of 3.1~eV from standard DFT. Oppositely, InMnO$_3$ has a calculated band gap of 1.4~eV, similar to YMnO$_3$, but a much smaller polarization \cite{makeref4SI}. This is also evident from the electrostatic potential profiles for the three systems in Fig.~\ref{fig:elstat_DFT}(a).

Using bulk polarizations from single unit cells (Table~\ref{tab:2}), and our total dielectric constants $\epsilon_{tot}$ (see Table~\ref{tab:1} and methods), the critical distance $d$ for electrostatic breakdown becomes 24.0~{\AA} (YMnO$_3$), 51.4~{\AA} (InMnO$_3$), and 45.1~{\AA} (YGaO$_3$), in qualitative agreement with Fig.~\ref{fig:charge_comp_Landau}. Thus, in order to render the DWs conducting in InMnO$_3$ and YGaO$_3$ by DFT modeling, larger supercell sizes are required (Fig. S5 for YGaO$_3$). Experimentally, the domain size can be controlled by the cooling rate through $T_C$~\cite{PhysRevX.2.041022,Meier.PhysRevX.7.041014}, hence both the scenarios described here, with DW distances smaller and larger than the critical distance $d$, can be achieved.\\

\textbf{Point defects and aliovalent dopants.} 
Above we showed a model where the inherent electric field is screened by the formation of electron-hole pairs. This screening mechanism, which is realized in defect-free lattices, comes with a high energetic cost of the order of the band gap. In the presence of defects or dopants, the situation is different since other screening mechanisms may be energetically more favorable. 

Hexagonal manganites are typically rendered $p$-type\cite{NatMater.9.253,NatCommun.7.13764}, caused by $R$ cation deficiency\cite{NatCommun.7.13764} during synthesis or incorporation of oxygen interstitials\cite{NatCommun.7.13745} during post-synthesis cooling. This gives asymmetric conductivity between the different walls, since the tail-to-tail walls will be screened by mobile holes\cite{NatMater.16.622}, while the band bending at the head-to-head walls remains for a finite screening length. We note that since Ga does not have Mn's multivalency, it is unknown if it intrinsically allows for oxygen interstitials. Hence, charge screening of the walls may be experimentally found to be fundamentally different in the gallates compared to the manganites. Oxygen vacancies would be expected to act as double electron donors, but there are to the best of our knowledge no experimental reports showing enhanced head-to-tail DW conductivity from oxygen vacancy formation. 

Even with equal concentration of field screening charge carriers, the conductivity of the head-to-head and tail-to-tail DWs may still differ due to carrier mobilities. Unlike most conventional semiconductors, electrons may be less mobile at head-to-head walls than holes are at tail-to-tail walls as electrons form polarons while holes are found in Bloch states at the respective walls \cite{NatMater.16.622}.

Aliovalent doping of h-$R$MnO$_3$ has been demonstrated to strongly modify the conductivity of charged DWs, without perturbating the DW pattern \cite{Schaab.AdvElectronMater.2.1500195,Hassanpour.NewJPhys.18.043015,Holstad.PhysRevB.97.085143}. Donor doping enhances the conductivity of head-to-head DWs, while acceptor doping promotes the conductivity of tail-to-tail DWs, fully in agreement with our presented model. Similar doping strategies may also enable conducting DWs in YGaO$_3$, InMnO$_3$ and other improper ferroelectrics.\\

\section{CONCLUSIONS}
In summary, we have calculated the energetics, crystal structure properties and electronic properties at charged ferroelectric DWs in YMnO$_3$, InMnO$_3$ and YGaO$_3$ by first principles calculations and phenomenological modeling. $\Delta\Phi=60^\circ$ DWs display lower formation energies than $\Delta\Phi=180^\circ$, in agreement with experiments. 

The similar DW widths in YMnO$_3$ and YGaO$_3$, and the wider DWs in InMnO$_3$, are correlated with the ferroelectric mode amplitude $Q$ and strength of the coupling term to $P$, reasoned from the ionic or covalent nature of the $R$-O$_p$ bond. 

Head-to-head and tail-to-tail DWs show asymmetric crystal structure behavior, which we attribute to the inherent difference in $R$ cation termination, and resulting local chemical environment, at the two walls.

Using a Zener-like electrostatic breakdown model, we determine the charge compensation, and resulting bulk polarization, for increasing DW distance. Bulk polarization is shown to be reduced for shorter wall distances. With the chosen DW distance, YMnO$_3$ shows charged DWs, in contrast with InMnO$_3$ and YGaO$_3$, evident from both electrostatic potential gradients and local electronic densities of states. This is explained by the inherent differences in polarization and electronic band gap of the three material systems. 

Through combined phenomenological model and first principles calculations, enhanced conductivity of charged DWs can be predicted based on DW distance (domain size), ferroelectric polarization and electronic band gap. 

\begin{acknowledgments}
We thank Yu Kumagai, Nicola A. Spaldin and Tsuyoshi Miyazaki for helpful discussions, and YK for also reading and commenting on our manuscript. D.R.S. acknowledges the Research Council of Norway (FRINATEK project no. 231430/F20), Norwegian University of Science and Technology (NTNU), and International Cooperative Graduate School program (ICGS) under the “Norwegian University of Science and Technology - NIMS Cooperative Graduate School program” fellowship for financial support. Q.M. acknowledges financial support by ETH Zurich and the Koerber foundation. K.I. acknowledges the Research Council of Norway (FRINATEK project no. 240466/F20). UNINETT Sigma2 - the National Infrastructure for High Performance Computing and Data Storage in Norway through projects ntnu243 and NN9264K, and ETH Zurich and by a grant from the Swiss National Supercomputing Centre (CSCS) under Project No. p504, are acknowledged for computing resources.
\end{acknowledgments}

\section*{METHODS}
\textbf{Density functional theory calculations.} DFT calculations were carried out with VASP \cite{PhysRevB.50.17953,PhysRevB.54.11169,PhysRevB.59.1758}, using the PBEsol functional\cite{PhysRevLett.100.136406}. $1\times1\times6$ supercells with one head-to-head and one tail-to-tail DW separated by three unit cells with a wall distance of $\sim$35~{\AA} were used as model systems. The plane wave energy cutoff was set to 550~eV and the Brillouin zone was sampled with a $\Gamma$-centered $4\times4\times1$ $k$-point grid for geometry optimization, and $6\times6\times1$ for density of states calculations. Lattice parameters were set to relaxed bulk values, and lattice positions relaxed until forces on the ions were below 0.005~eV{\AA}$^{-1}$. GGA+U\cite{PhysRevB.57.1505} with U = 5~eV on Mn $3d$ reproduced the experimental band gap\cite{ApplPhysB.73.139} and lattice parameters\cite{PhysRevB.83.094111,PhysRevB.85.174422,ActaCrystallogrB.31.2770}. No U was applied for YGaO$_3$ (see Fig S6 \cite{makeref4SI}). YMnO$_3$ and InMnO$_3$ were initialized with collinear frustrated antiferromagnetic order\cite{JPhysCondensMatter.12.4947} on the Mn sublattice.

InMnO$_3$ supercells were relaxed in two steps, initially with a force criterion for ions of 0.04~eV{\AA}$^{-1}$, and finally with 0.005~eV{\AA}$^{-1}$, where $\Phi$ and $\alpha_A$ at the domain centers were locked (Fig. S7 \cite{makeref4SI}).

DW formation energies were calculated as
\begin{equation}\label{eq:Efdw}
E^f_{DW}=\dfrac{1}{2A} \left(E^f_{DW~struct.}-E^f_{ref}\right) \quad ,
\end{equation} 	
where $E^f_{DW~struct.}$ is the total energy of the supercell with two DWs and cross-sectional area $A$, and $E^f_{ref}$ the energy of the monodomain supercell. Both DWs are assumed to contribute eq	ually to the total energy of the system.\\

\textbf{Landau coefficients.} Landau Free Energy parameters were calculated using DFT as implemented in abinit \cite{Gonze:2002br,Gonze:2005em,Torrent:2008jw}.
Frozen phonon calculations on the high-symmetry $P6_3/mmc$ unit cell were done with a 30 atom supercell. The force constant matrix was extracted with phonopy \cite{phonopy}.
LDA+U \cite{Amadon:2008ia} with a U = 8~eV and J = 0.88 on Mn 3$d$ was used for YMnO$_3$ and InMnO$_3$. The plane wave cutoff was set to 30~Ha, and $k$-point grids of $6\times6\times2$ and $4\times4\times2$ were used for the high and low symmetry unit cells, respectively. 
Different eigenvector amplitudes of the force constant matrix were superimposed and fitted to the free energy functional \cite{NatMater.13.42}. The gradient terms were extracted from the dispersion of the $K_3$ branch of the force-constant matrix. The calculated values are tabulated in the Supplementary material \cite{makeref4SI}.\\

\textbf{Minimization of the Landau Free Energy.} DW widths and amplitudes of the different modes were calculated by minimizing the Landau Free Energy with fixed boundary conditions. A 200~{\AA} grid with a grid width of 1~{\AA} was used and the grid size was tested for convergence. A constrained BFGS-alghoritm\cite{BFGS.Flet87} was applied to minimize the functional
\begin{equation*}
	f=\int dx F[Q,\Phi,P]
\end{equation*}
over the whole grid.

Electrostatic minimizations were performed starting with the the non-electrostatic minimization and minimizing the polar mode under constant trimerization. Proper minimization could not be done due to attractive forces between the walls, and because reducing the polarization leads to increasing degeneracy between the values of $\Phi$, DW broadening and poor convergence.

Charge compensation of the DWs in the Landau model was calculated by introducing electron-hole pairs at the walls at the cost of the band gap energy $E_g$. The spontaneous polarization was then self-consistently minimized while accounting for charge transfer between the walls as well as other screening effects.\\

\textbf{Dielectric constants.} The total energy stored in the electromagnetic fields can be expressed as:
\begin{equation}
F(P)=-\dfrac{1}{2}\int d^3x \epsilon_0\epsilon_b\mathcal{E}^2-\int d^3x\vec{P}\cdot\vec{\mathcal{E}} \quad ,	
\end{equation}
where $P$ is the spontaneous polarization and $\mathcal{E}$ is the internal electric field. Here the background dielectric constant $\epsilon_b$ contains the response of all the normal modes in the system except the one corresponding to $P$.

Under a static field $\epsilon_b$ can be expressed as:
\begin{equation}
\epsilon_b=1+\chi_{\infty}+\sum\limits_{n\neq n_{P}}\chi_n\quad ,
\end{equation}
where $\chi_\infty$ is the electronic response and
\begin{equation}
\chi^{-1}_n=\epsilon_0\dfrac{a_n}{Z_n^2}\Omega \quad .
\end{equation}
Here $a_n$ are the eigenvalues of the force constant matrix and $Z_n$ is the effective charge of the modes. These values were extracted using DFPT as implemented in abinit \cite{Gonze1997a}. The three compounds were found to exhibit two more displacements with non-zero effective charge $Z$ in the $z$-direction in addition to the ferroelectric mode, leading to total background dielectric constant $\epsilon_b$ of 8.9 (YMnO$_3$), 6.7 (YGaO$_3$) and 10.3 (InMnO$_3$).
We can estimate the full dielectric constant from the high-symmetry structure by realizing that the the expectation value of the spontaneous polarization is given by:
\begin{equation}
P(\mathcal E)=\dfrac{\mathcal E}{g'\left<Q\right>^2+a_p} + \dfrac{gQ^3}{g'\left<Q\right>^2+a_p} \quad ,
\end{equation}
where the second part is the spontaneous polarization and the first part is induced by the field, thus we find that the term containing g' adds additional stiffness to the polar mode.
Which means that by choosing the appropriate units we can calculate the static susceptibility of the polar mode ($\chi_P$) by: 
\begin{equation}
\chi_P^{-1}=\epsilon_0(g'\left<Q\right>^2+a_p) \quad .
\end{equation}
Adding these terms to the background dielectric constant we find a total dielectric  constant $\epsilon_{tot}$ of 12.5 (YMnO$_3$), 11.2 (YGaO$_3$) and 14.8 (InMnO$_3$). 

\bibliography{references}

\end{document}